\documentclass[12pt]{iopart}
\usepackage{subfigure}
\usepackage{fullpage}
\usepackage{setspace}
\usepackage{parskip}
\usepackage{titlesec}
\usepackage[section]{placeins}
\usepackage{xcolor}
\usepackage{breakcites}
\usepackage{lineno}
\usepackage{hyphenat}
\usepackage{graphicx}
\usepackage[space]{grffile}
\usepackage{latexsym}
\usepackage{textcomp}
\usepackage{longtable}
\usepackage{tabulary}
\usepackage{booktabs,array,multirow}
\expandafter\let\csname equation*\endcsname\relax
\expandafter\let\csname endequation*\endcsname\relax
\usepackage{amsfonts,amsmath,amssymb}

\newtheorem{thm}{Theorem}[section]

\newtheorem{lemma}[thm]{Lemma}
\newtheorem{corollary}[thm]{Corollary}
\newtheorem{proposition}[thm]{Proposition}

\newtheorem{defn-thm}[thm]{Definition-Theorem}

\begin{document}

\title[Author guidelines for IOP Publishing journals in  \LaTeXe]{Witten-type topological field theory of self-organized criticality for stochastic neural networks }

\author{Jian Zhai$^1$, Chaojun Yu$^1$, You Zhai$^2$}

\address{1: School of Mathematical Sciences, Zhejiang University, Hangzhou, China}
\address{2: The First Affiliated Hospital, Zhejiang University, Hangzhou, China}
\ead{jzhai@zju.edu.cn}
\vspace{10pt}
\begin{indented}
\item[]May 2021
\end{indented}

\begin{abstract}
We study the Witten-type topological field theory(W-TFT) of self-organized criticality(SOC) for stochastic neural networks.  The Parisi-Sourlas-Wu quantization of general  stochastic differential equations (SDEs) for neural networks,  the Becchi-Rouet-Stora-Tyutin(BRST)-symmetry of the diffusion system and the relation between spontaneous breaking and instantons connecting steady states of the SDEs, as well as the sufficient and  necessary  condition on pseudo-supersymmetric  stochastic neural networks are obtained. Suppose  neuronal avalanche is a mechanism of cortical information processing and storage \cite{Beggs}\cite{Plenz1}\cite{Plenz2}  and the model of stochastic neural networks\cite{Dayan} is correct, as well as the SOC system can be looked upon as a W-TFT with spontaneously broken BRST symmetry. Then we should recover the neuronal avalanches and spontaneously broken BRST symmetry from the  model of stochastic neural networks.   We  find that, provided the divergence of drift coefficients is small and non-constant, the model of stochastic neural networks is BRST symmetric. That is, if the SOC of brain neural networks system can be looked upon as a W-TFT with spontaneously broken BRST symmetry, then the general model of stochastic neural networks which be extensively used in neuroscience \cite{Dayan} is not enough to describe the SOC. On the other hand, using the Fokker-Planck equation, we show the sufficient condition on diffusion so that there exists a steady state probability distribution for the stochastic neural networks. Rhythms of the firing rates of the neuronal networks  arise from the process,  meanwhile some biological laws are conserved.

\end{abstract}

%
%
%
%
%
\maketitle

\section {introduction}

It is well known that brain neural networks are organized based on self-organized criticality(SOC). The power-law of background neural activities  had been discovered in EEG, ECoG and fMRI (see, for example, \cite{Buzsaki}\cite{FZ}\cite{He} \cite{Beggs}etc.). Beggs and Plenz \cite{Beggs}\cite{Plenz1}\cite{Plenz2} studied spontaneous activity in an isolated slab of cortical tissue which followed up a much earlier series of experiments on isolated cortical slabs or slices carried out by Burns \cite{Burns}. Their work revealed that an isolated cortical slab remains silent but excitable by brief current pulses. A strong enough pulse can trigger a sustained
all-or-none response that propagates radially from the stimulation site, at a velocity of
about $15 (cm/s)$. Propagation of local field potentials(LFPs) in cortical circuits could be described by the same power-law that govern avalanches.

Can neuronal avalanches and  SOC of brain arise from current often utilized general  mathematical models for brain neural networks?  Cowan, Neuman, Kiewiet and  Drongelen \cite{Cowan} analyzed  neural networks that
exhibits self-organized criticality. Such criticality follows from the combination
of a simple neural network with an excitatory feedback loop that generates
bistability, in combination with an anti-Hebbian synapse in its input pathway.
Using the methods of statistical field theory, they show that
the network exhibits hysteresis in switching back and forward between its two
stable states, each of which loses its stability at a saddle{node bifurcation. The Biological conservation law as an emerging functionality in dynamical neuronal networks was studied in \cite{Podobnik}. However, so many different models were utilized in different papers.  The lack of unifying mathematical theory and analyzing methods for the SOC impedes the effort to understand the neural information process arisen in our brain.

 In \cite{Ovchi}, a scenario that a generic SOC system can be looked as a Witten-type topological field theory (W-TFT) with spontaneously broken Becchi-Rouet-Stora-Tyutin(BRST) symmetry, was proposed. But there are at least two unsolved problems: (1) From the quantum mechanical treatment, it is not clear why the distribution of avalanches must be a power-law; (2) Suppose we identify the instanton-induced $\mathcal{Q}$-symmetry breakdown, when the spontaneous breakdown of $\mathcal{Q}$ symmetry occurs.

In this paper, we consider a general standard model of stochastic neural networks which was extensively used in neuroscience \cite{Dayan}. Suppose  the neuronal avalanches is a mechanism of cortical information processing and storage \cite{Beggs}\cite{Plenz1}\cite{Plenz2}  and the model of the stochastic neural networks\cite{Dayan} is correct, as well as the SOC system can be looked upon as a Witten-type topological field theory (W-TFT) with spontaneously broken Becchi-Rouet-Stora-Tyutin
(BRST) symmetry. Then we should recover  Beggs and Plenz etc.'s results and find power-law as well as spontaneously broken BRST symmetry from the  model of stochastic neural networks.   We utilize Parisi-Sourlas-Wu quantization to the stochastic differential equations (SDEs) of the standard model of stochastic neural networks and find that, provided the divergence of drift coefficients is small and non-constant, the model of stochastic neural networks is  BRST symmetric in the sense of W-TFT. That is, if the SOC of brain neural networks system can be looked upon as a W-TFT with spontaneously broken BRST symmetry, then the general model of the stochastic neural networks which be extensively used in neuroscience \cite{Dayan} is not enough to describe the SOC. At the same time, we solve the stochastic differential equations for the model  of stochastic neural networks by numerical method, and try to recover Beggs and Plenz etc.'s results and find power-law. On the other hand, we study the sufficient condition on diffusion such that there exists a stationary probability distribution for the stochastic neural networks. The diffusion processes (solutions to the SDEs) take a long time  inside some  maximal equivalent class of the unperturbed diffusion process, and suddenly departing  from one maximal equivalent class (up-state or down-state) and arriving to another (down-state or up-state).  Rhythms of the firing rates of the neuronal network  arise from the transitions  between the maximal equivalent classes (tunneling effect),  meanwhile some biological laws are conserved.

\vspace{0.5cm}

\section {stochastic neural networks}

Consider a neural network of $N$ neurons where each neuron receiving $N$ synaptic inputs labeled by $i=1,2,...,N$. The firing rate of input $i$ is denoted by $v^i$ and the input rates are represented collectively by the $N$-component vector $v$. The synaptic current $I_s^i$ of $i$th neuron is generally modeled by (see \cite{Dayan} section 7.2 (7.6))
\begin{equation}\label{10.1}
\tau_s dI_s^i=\left(-I_s^i-E^i(u(t))\right)dt+\tau_s\alpha_c^i(I_s(t),t)d\eta^i(t),
\end{equation}
where $\tau_s$ is the decay constant of synaptic conductance, $E^i(u):=\partial_{u^i}E(u)$, $\{\eta^i(t)\}_i$ are independent Wiener processes and $\alpha_c^i$ are diffusion coefficients. For simplicity, assume $\alpha_c^i(I_s(t),t)=\alpha_c^i(I_s(t))$.

For constant synaptic current, the firing rate of postsynaptic neuron $i$ can be expressed as $F(I_s^i)$ that will be used as the input firing rate at next time, where $F$ is an increasing function and called an activation function. Assume that the time-dependent inputs are still given by this activation function
\begin{equation}\label{10.4}
u^i(t)=F(I_s^i(t)).
\end{equation}
$F$ is sometimes taken to be a saturating function such as a sigmoid function. It is also bounded from above, which can be important in stabilizing a network against excessively high firing rates.

\vspace{0.5cm}

\subsection{ Stratonovich interpretation }

By Ito formula, (\ref{10.1})(\ref{10.4}) imply
\begin{equation}\label{10.5}\begin{aligned}
du^i =\left(-F'(I_s^i)\frac{I_s^i+E^i(u)}{\tau_s}+\frac12 F''(I_s^i)(\alpha_{c}^i(I_s))^2\right)dt+F'(I_s^i)\alpha_{c}^i(I_s) \circ d\eta^i
\end{aligned}\end{equation}
where, from the view point of self-organized criticality (SOC)\cite{Ovchi}, we use the Stratonovich interpretation of solutions to SDE (\ref{10.5}). Suppose $u^i$ is the Stratonovich integration of SDE (\ref{10.5}). The Ito's equivalent SDE of $u^i$
\begin{equation}\label{10.7}
du^i =\left(-F'(I_s^i)\frac{I_s^i+E^i(u)}{\tau_s}+\frac12 F''(I_s^i)(\alpha_{c}^i(I_s))^2+\frac14\frac{\partial}{\partial u^i}(F'(I_s^i)\alpha_{c}^i(I_s))^2\right)dt+F'(I_s^i)\alpha_{c}^i(I_s) d\eta^i.
\end{equation}

\vspace{0.5cm}

\subsection{sufficient condition on the existence of  stationary probability distribution}

\begin{lemma}\label{L10.1}
If $\alpha^i_{c_0}(I_s(t)):=\lim_{c\to c_0}\alpha_c^i(I_s(t),t)=\sqrt{2T/F'(I_s^i(t))}$, then there is a  stationary probability distribution of (\ref{10.7})    with the following stationary density
$$
P_0(u)=\frac1{Z}\exp(\frac{-1}{T\tau_s}\tilde{E}(u))
$$
where  $\tilde{E}(u):=\frac1{\tau_s}(\sum_i\int^{u^i}\left(F^{-1}(r)-\frac{\tau_sT f'(r)}{2f(r)} \right)dr+E(u))$, $f(r)=F'(F^{-1}(r))$ and
$$
\quad Z=\int_u[du] \exp(\frac{-1}{T\tau_s}\tilde{E}(u)).
$$
\end{lemma}

$Proof$. As $c\to c_0$, (\ref{10.7}) is written as
\begin{equation}\label{10.9}\begin{aligned}
du^i =\left(-f(u^i)\tilde{E}^i(u)+T f'(u^i)\right)dt+\sqrt{2Tf(u^i)}d\eta^i.
\end{aligned}\end{equation}

Generally, suppose $u=\{u_i\}_i$ satisfy following stochastic differential equations
$$
du_i(t)=b_i(u(t))dt+\sigma_i(u(t))d\eta^i(t).
$$
The transition density $P(u,t | u(t_0),t_0)$ of $u(t)$ must satisfy the Fokker-Planck equation
$$
\frac{\partial P}{\partial t}=\sum_i\frac{\partial}{\partial u^i}\left(\frac12\frac{\partial}{\partial u^i}(\sigma_i^2P)-b_iP   \right).
$$
It follows that if
$$
b_i=\frac12\sigma_i^2\frac{-\tilde{E}^i(u)  }{ T }+\frac{\partial}{\partial u^i}\frac{\sigma_i^2}2
$$
 then $P_0=\frac1{Z}\exp(\frac{-1}{T}\tilde{E}(u))$ is stationary to the Fokker-Planck equation. Note that
\begin{equation}\label{10.6.1}
 \tilde{E}(u)(t)=\tilde{E}(u)(0),\quad\forall t
 \end{equation}
 is conserved.

Remark.  Recall that the drift part of (\ref{10.9}) is proportional to the gradient flow of the energy $\tilde{E}(u)$.  As the parameter $T$ (temperature) decreases to zero, $P_0(u)$ will approach a stationary transition density on a set of minima of $\tilde{E}(u)$. This is the principle on which simulated annealing is based \cite{KGV}. For the study of the behavior of the solutions to the SDE on large time intervals for small noise, an essential role is placed by the Markov chain on the set of boundaries of $\omega$-limit sets (see Fig.4). The stationary probability distribution of the diffusion process determined by SDE can be described by the stationary probability distribution of the Markov chain.   The diffusion processes (solutions to the SDEs (\ref{10.9})) take a long time  inside some  maximal equivalent class of the unperturbed diffusion process, and suddenly departing  from one maximal equivalent class (up-state / down-state) and arriving to another (down-state / up-state).  Rhythms of the firing rates of the neuronal network  arise from the transitions  between the maximal equivalent classes of steady states,  which is an important characteristic of brain neuronal networks.

\vspace{0.5cm}

 As $c\to c_0$, the Stratonovich interpretation of  SDE (\ref{10.5}) is rewritten as
\begin{equation}\label{10.6}\begin{aligned}
du^i =\left(-f(u^i)\hat{E}^i(u)+T f'(u^i)\right)dt+\sqrt{2Tf(u^i)}\circ d\eta^i, \quad \hat{E}(u):=\frac1{\tau_s}(\sum_i\int^{u^i}(F^{-1}(r)dr+E(u)).
\end{aligned}\end{equation}
Since the diffusion term $\sqrt{2Tf(u^i)}$ in (\ref{10.6}) depending on $u$, to avoid the quantization accounting the curvature of the target manifold,  take the transformation
$$
\Phi(r):=\int_0^r\frac1{\sqrt{f(s)}}ds
$$
and $x^i=\Phi(u^i)$. Then
\begin{equation}\label{10.10.0}
dx^i=\left(\frac{-f(u^i)\hat{E}^i(u)+Tf'(u^i)}{\sqrt{f(u^i)}  }\right)dt+\sqrt{2T}\circ d\eta^i=:A^i(x)dt+\sqrt{2T}\circ d\eta^i
\end{equation}
is deduced from (\ref{10.6}). Its Ito's equivalent SDE
\begin{equation}\label{10.10}
dx^i=\left(\frac{-f(u^i)\tilde{E}^i(u)+Tf'(u^i)}{\sqrt{f(u^i)}  }\right)dt+\sqrt{2T} d\eta^i.
\end{equation}
 Remark that
\begin{equation}\label{10.11}
\partial_{x^j}A^i=\partial_{x^i}A^j,
\end{equation}
and there is a potential function $V(x)$ such that $A$ can be rewritten via  $V$:
\begin{equation}\label{10.12}
A^i(x)=\partial_{x^i}V(x).
\end{equation}

\vspace{0.5cm}

\section{ quantization  }

 Generally (\ref{10.5}) is rewritten as an $N$-dimensional diffusion process
\begin{equation}\label{6.2}
dv^{i}=A^i(v)dt+\sigma^i(v)\circ d\eta_t^i,\quad i=1,2,...,N
\end{equation}
where $\eta_t$ is the $N$-dimensional Wiener process, and $A(v)=\nabla_v V(v)+\tilde{A}(v)$. The nonpotential part $\tilde{A}$ is regarded as a magnetic field.

Following \cite{Ovchi}\cite{Ovchi_1}, where SOC was interpretation as Witten-type topological field theory with spontaneously broken BRST symmetry \cite{BBRT}. To complete this theory,  there are at least two problems need to be fixed. First, the BRST-symmetry breakdown by instantons is only proved in one-dimension (\cite{BBRT}p.170). In multiple dimension case, even without the magnetic part $\tilde{A}$, the nonzero average $\mathbb{Q}_{\alpha\beta}$ of a $\mathcal{Q}$-exact operator only implies the perturbative ground state $\langle\alpha |$ doesn't determine a supersymmetric ground state in the full theory (\cite{Mirror}p.223). It is unclear when a Hamiltonian related to $A$ is pseudo-Hermitian and when the hopping evolution of instanton and anti-instanton along the magnetic field  $\tilde{A}$ induces pseudo-supersymmetry breakdown. Second, it is not clear why the BRST breakdown must be a power-law.

\subsection{Quantization by the De Witt-Faddeev-Popov method}
We proceed with  quantization \cite{PSW} of equation (\ref{6.2}) by the De Witt-Faddeev-Popov method. In light of the equation (\ref{10.10}), the path integral representation of the Witten index is
\begin{equation}\label{10.13}
\mathcal{Z}=\int_{\xi}[d\xi] e^{-\frac12\int_0^{\tau} g_{jk}\xi^j\xi^k dt}\frac{det(\frac{\delta\xi}{\delta v})}{|det(\frac{\delta\xi}{\delta v})|}=\int_{v}[dv] e^{-\frac12\int_0^{\tau} g_{jk}(\partial_tv^j-A^j)(\partial_tv^k-A^k) dt}det(\frac{\delta\xi}{\delta v})
\end{equation}
$g^{jk}=(g_{jk})^{-1}$ is the noise-noise correlator
\begin{equation}\nonumber
\langle\xi^j(t)\xi^k(t')\rangle=g^{jk}\delta(t-t'),\quad d\xi^j(t):=\sigma^j(v) d\eta^j
\end{equation}
which define a metric on the target manifold $M\ni v$ and which for now is assumed to be independent of the field $v$, and  $det(\frac{\delta\xi}{\delta v})$ is the Jacobian of the map defined by the equation (\ref{10.5}). Suppose $ \sigma^j(v)\equiv 1 $ (similar conclusions can be obtained for general case in the same spirit). By Gaussian multiple integral, the Jacobian can be represented as the path integral over the fermionic Faddeev-Popov ghosts
\begin{equation}\nonumber
\tau det(\frac{\delta\xi}{\delta v})=\int [d\psi] [d\bar\psi] e^{i\int_0^{\tau}\bar\psi_j(\delta^j_k\partial_t-\partial_{v^k}A^j)\psi^kdt}(\int[d\psi] [d\bar\psi']e^{i\int_0^{\tau}\bar\psi_j'\psi^jdt})^{-1}
\end{equation}
and  the path integral representation of the Witten index now is
\begin{equation}\label{10.14}\begin{aligned}
\mathcal{Z}&=\int_{v,\psi,\bar\psi} e^{-\frac12\int_0^{\tau} g_{jk}(\partial_tv^j-A^j)(\partial_tv^k-A^k) -i\bar\psi_j(\delta^j_k\partial_t-\partial_{v^k}A^j)\psi^k dt}\\
&=(det \mathcal{J})^{-1}\int_{v,B,\psi,\bar\psi} e^{-\int_0^{\tau} i g_{jk}(\partial_tv^j-A^j)B^k+\frac12g_{jk}B^jB^k -i\bar\psi_j(\delta^j_k\partial_t-\partial_{v^k}A^j)\psi^k dt}
\end{aligned}\end{equation}
where the last equality derived by the Faddeev-Popov method.

\subsection{BRST symmetry}
The path integral representation of the Witten index is invariant under the nilpotent infinitesimal BRST transformation
\begin{equation}\label{10.15}
\{Q,v^j\}=\psi^j,\quad \{Q, \psi^j\}=0,\quad \{Q, \bar\psi_j\}=B_j,\quad \{Q, B_j\}=0,\quad \{Q,Q\}=0
\end{equation}
and the action is $Q$-exact
\begin{equation}\label{10.16}
\int_0^{\tau} ig_{jk}(\partial_tv^j-A^j)B^k+\frac12g_{jk}B^jB^k -i\bar\psi_j(\delta^j_k\partial_t-\partial_{v^k}A^j)\psi^k dt=\int_0^{\tau} \{Q,\bar\psi_j(i(\partial_tv^j-A^j)+\frac12 B^j)\}dt
\end{equation}
where $\{\cdot,\cdot\}$ is Poisson brackets (see \cite{BBRT}(3.65)), $Q=-i\psi^jB^j$ is the BRST operator.

Suppose $\partial_{v^j}A^k=\partial_{v^k}A^j$. As \cite{BBRT}(3.62),  the action
$$
\int_0^{\tau} ig_{jk}(\partial_tv^j-\sigma A^j)B^k+\frac12g_{jk}B^jB^k -i\bar\psi_j(\delta^j_k\partial_t-\sigma\partial_{v^k}A^j)\psi^k dt
$$
is invariant under the discrete transformation
\begin{equation}\label{10.18}
\sigma\to-\sigma,\quad \psi\to\bar\psi,\quad \bar\psi\to\psi,\quad B\to B-2i\sigma A
\end{equation}
and there is a second BRST operator $\bar Q:=-i\bar\psi^k(B^k-2iA^k)$ with $\sigma=1$.
 It is remarkable that, even if $\partial_{v^j}A^k\neq\partial_{v^k}A^j$, utilizing the Poisson brackets we still can prove that  $\bar Q$ is  nilpotent and
\begin{equation}\label{10.19.1}
 \{Q,\bar Q\}=2i(\frac12(\partial_tv^j-A^j)^2+(\partial_tv^j-A^j)A^j-i\bar\psi^j\partial_{v^k}A^j\psi^k   )=2i H.
\end{equation}
The interaction representations of the BRST operators
\begin{equation}\label{10.20}\begin{aligned}
&Q=-i\psi^jB^j=-\psi^j(\partial_tv^j-A^j)\to \mathcal{Q}=-\psi^j\partial_{v^j},\\
&\bar Q=-i\bar\psi^k(B^k-2iA^k)=-\bar\psi^k(\partial_tv^k+A^k)\to\bar{\mathcal{Q}}=-\bar\psi^k(\partial_{v^k}+2A^k)
\end{aligned}\end{equation}
and the Hamiltonian
\begin{equation}\label{10.21}
H=\frac{-1}2\Delta_v-A\cdot\nabla_v-div_v(A),
\end{equation}
are derived by an appropriate quantum
\begin{equation}\label{10.21.1}\begin{aligned}
&\Pi^j:=\frac{\delta^RL}{\delta v^j}=\partial_tv^j-A^j\to \partial_{v^j},\\
&\Pi^jA^j\to \frac12 [\partial_{v^j},A^j]_{+}=A\cdot\nabla_v+\frac12div_vA,\\
& -i\bar{\psi}^j\psi^k\to\frac12[\partial_{\psi^k},\psi^j]_{-}=\frac12\delta^{jk}.
\end{aligned}\end{equation}
Notice that
\begin{equation}\label{10.21.2}
\mathcal{Q}\bar{\mathcal{Q}}=i(\frac{-1}2\Delta_v-A\cdot\nabla_v-div_v(A))=iH.
\end{equation}

The natural representation of the above algebra of observable operators is on the space of differential forms (see \cite{Mirror}(10.214)(10.225))
$$
\mathcal{H}=\Omega(M)\bigotimes \mathbb{C}
$$
equipped with the Hermitian inner product
$$
(\omega_1,\omega_2)=\int_M \bar\omega_1 \wedge \ast\omega_2.
$$

\subsection{pseudo-supersymmetry}
Generally \cite{Mostafaza}, $H$ is  pseudo-Hermitian if there is a linear invertible Hermitian $\Lambda$ and the adjoint operator of $H$ satisfies
$$
H^\dag=-\Lambda H\Lambda^{-1}.
$$
If $\mathcal{Q}^\sharp:=\Lambda^{-1} \mathcal{Q}^\dag\Lambda$ is also nilpotent  and $2iH=\{Q,Q^\sharp\}$, we call the system $2$-pseudo-supersymmetric \cite{Mostafaza}. Here the notations are slightly different from \cite{Mostafaza} where the Hamiltonian $H_M$ utilized in \cite{Mostafaza} satisfies $2H_M=\{Q,Q^\sharp\}$. So we have to use $H_M:=iH$. Furthermore, if $\mathcal{Q}$ is BRST and there exists at least one physical state in the sense of gauge theory or W-TFT \cite{BBRT}, the system is called BRST-supersymmetric in the sense of gauge theory or W-TFT respectively. Otherwise, the supersymmetries are broken. The BRST-supersymmetry breaking is spontaneous.

In light of (\ref{10.21.2}), $H$ is  pseudo-Hermitian with respect to $\Lambda$ if and only if
$$
-i\mathcal{Q}\bar{\mathcal{Q}}=H=-\Lambda^{-1}H^\dag\Lambda=-i\Lambda^{-1}\bar{\mathcal{Q}}^\dag\Lambda\Lambda^{-1}\mathcal{Q}^\dag\Lambda.
$$

\begin{proposition}\label{th.10.2}
The system (\ref{6.2}) is  2-pseudo-supersymmetric if  there is a linear invertible Hermitian $\Lambda$ s.t.
$$
\bar{\mathcal{Q}}=\Lambda^{-1}\mathcal{Q}^\dag\Lambda.
$$
\end{proposition}

\begin{corollary}\label{co.10.1}
If $A$ is given via a potential $V$: $A=\nabla_v V$, then the system (\ref{6.2}) is  2-pseudo-supersymmetric.
\end{corollary}

$Proof.$ Take $\Lambda=\exp\{2V\}$ and note that
$$
\mathcal{Q}^\dag\Lambda=\Lambda(-\bar\psi)(\nabla_v+2A)=\Lambda \bar{\mathcal{Q}}  .
$$

\subsection{spectrum}

 It was proved in \cite{Mostafaza} that the Hamiltonian $H$ is pseudo-Hermitian if the spectrum of $H$ consists of real and the pairs of complex-conjugate energies. It was well known that, when the Hamiltonian is self-adjoint(i.e. Hermitian), all eigenvalues must be real. So non-self-adjoint  is a necessary condition of pseudo-supersymmetric breaking. It is easy to see that

\begin{lemma}\label{L10.3}
$H$ is self-adjoint(i.e. Hermitian) in $L^2$ if and only if
$$
A(v)\equiv 0.
$$
Moreover $A(v)=0$ implies that $v$ is a steady state of the unperturbed  dynamical system of the equation (\ref{6.2}).
\end{lemma}

\vspace{0.5cm}

Consider the spectrum of the Hamilton $H$ on $M$  with periodic boundary condition(P.B.C.) or Dirichlet  boundary condition(D.B.C.) or Neuman  boundary condition(N.B.C.).

\begin{thm}\label{th.10.4} $(\sigma I+H)^{-1}$ for some $\sigma>0$ is compact, positive  from $W^{1,2}(M)$ to itself. Moreover, suppose any two eigenvectors are orthogonal. Then all the eigenvalues of $H$ are real if and only if $A(v)\equiv 0.$
\end{thm}

$Proof.$ The spectrum problem of the Hamilton $H$ is equivalent to
\begin{eqnarray*}
H_\sigma\Psi:=(\sigma I+H)\Psi=(\lambda+\sigma)\Psi
\end{eqnarray*}
that can be extended to a weak form in $W^{1,2}(M)$
\begin{eqnarray*}
{\mathcal H}_{\sigma}v-\sigma E_1E\Psi=\lambda E_1E\Psi,
\end{eqnarray*}
where ${\mathcal H}_{\sigma}$ is the extension of $H_{\sigma}$ on $W^{1,2}(M)$, $E:  W^{1,2} \to  L^2$ and $E_1: L^2 \to  (W^{1,2})^\ast$ are two embedding maps. From the Sobolev theorem, the embedding map $E: W^{1,2}\to  L^2$ is compact. while $E_1:  L^2 \to (W^{1,2})^\ast$ is continuous. So ${\mathcal H}_{\sigma}^{-1}E_1E: W^{1,2}\rightarrow W^{1,2}$ is compact. Similar proof can be applied to D.B.C. or N.B.C..

To study complex eigenvalue of $H$, we have to consider the Hilbert space  $W^{1,2}(M)$ of complex valued functions.  Then $H$ is self-adjoint(i.e. Hermitian) if and only if all the eigenvalues are real. By Lemma \ref{L10.3}, all the eigenvalues of $H$ are real if and only if $A(v)\equiv 0.$

\subsection{BRST supersymmetric breaking}
It is easy to see that $H$ is pseudo-Hermitian.  First we are looking for eigenvalues
\begin{equation}\label{10.26.0}
H_0\Phi:=\frac{-1}2\Delta_x\Phi-A^j(x)\nabla_{x^j}\Phi=\lambda\Phi, \quad \forall x\in M
\end{equation}
with P.B.C or D.B.C or N.B.C. It is easy to see that, in case of P.B.C. or N.B.C., any non-zero constant is an eigenvector with zero eigenvalue.  Generally we have

\begin{lemma}\label{L3.5} Suppose \\
either (1)  $div_xA(x)=0$ and $\Phi$ satisfies D.B.C  or $div_xA(x)=0$ and $\Phi$ as well as $A$ satisfy P.B.C.;\\
or (2) $\Phi$ satisfies D.B.C., $M$ is star shape, $\max_{x\in M}|x||A(x)|<\frac{N-2}4$ and $|div_xA(x)|< C_M$ where $C_M$ is the Poincar\'{e} constant for D.B.C.;\\
or (3) $\Phi$ and $A$ satisfy P.B.C., $|\nabla_x\Phi(x)|^2\leq 2|\partial_\nu\Phi(x)|^2$($\forall x\in\partial M$), $M$ is symmetric under the transformation $x_j \to -x_j$ ($\forall j=1,2,...,N$), $\max_{x\in M}|x||A(x)|<\frac{N-2}4$ and $|div_xA(x)|< C_M$ where $C_M$ is the Poincar\'{e} constant for P.B.C.;\\
or (4) there is $V$ such that $A=\nabla_x V$, and $\Phi$ satisfies D.B.C. or N.B.C. or $\Phi$ and $V$ satisfy P.B.C..\\
  Then, in case of P.B.C. or N.B.C.  the only lowest energy state of  $H_0$ is a constant with zero eigenvalue, and in case of D.B.C. the lowest energy state of  $H_0$ is positive.
\end{lemma}

$Proof.$ (1) Note that
\begin{equation}\label{10.27}\begin{aligned}
\frac12\int_{M}|\nabla_x\Phi(x)|^2dx=Re(\lambda)\int_{M}|\Phi(x)|^2dx
\end{aligned}\end{equation}
which implies that the only lowest energy state is a constant with zero eigenvalue.

(2)(3) First note that the symmetry of $M$  under the transformation $x_j \to -x_j$ ($\forall j=1,2,...,N$) implies that $M$ is star shape. Multiply $H_0\Phi=\lambda\Phi$ by $\bar\Phi-\frac1{|M|}\int_M\bar\Phi dx$ and integrate over $M$
\begin{equation}\label{10.28.0}\begin{aligned}
&\frac12\int_{M}|\nabla_x(\Phi(x)-\frac1{|M|}\int_M\Phi )|^2dx+\frac12\int_M divA(x)|\Phi(x)-\frac1{|M|}\int_M\Phi |^2dx\\
&=Re(\lambda)(\int_{M}|\Phi(x)|^2dx-\frac1{|M|}|\int_M\Phi(x)dx|^2 ),\quad\forall \Phi\in W^{1,2}(M)\quad \text{with P.B.C.}
\end{aligned}\end{equation}
and recall the Schwartz inequality and Poincar\'{e} inequality
\begin{equation}\label{10.28.1}\begin{aligned}
&\int_{M}|\Phi(x)|^2dx-\frac1{|M|}|\int_M\Phi(x)dx|^2\geq0,\\
&\int_{M}|\nabla_x(\Phi(x)-\frac1{|M|}\int_M\Phi )|^2dx\geq C_M\int_{M}|\Phi(x)-\frac1{|M|}\int_M\Phi |^2dx> \int_M divA(x)|\Phi(x)-\frac1{|M|}\int_M\Phi |^2dx.
\end{aligned}\end{equation}
We discover that the eigenvalues of $H_0$ are non-negative.
 Furthermore, multiply $H_0\Phi=0$ by $x\cdot\nabla_x\bar\Phi$ and integrate over $M$
\begin{equation}\label{10.28}\begin{aligned}
&\frac{N-2}4\int_{M}|\nabla_x\Phi(x)|^2dx+\int_{\partial M}\left(\frac12Re\left(\partial_\nu\Phi(x)x\cdot\nabla_x\bar\Phi(x)\right)-\frac14 (x\cdot\nu(x)) |\nabla_x\Phi(x)|^2 \right)d\mathcal{H}^{N-1}(x)\\
&=-Re\int_{M}(A(x)\cdot\nabla_x\Phi(x))(x\cdot\nabla_x\bar\Phi(x))dx\\
&\leq (\max_{x\in M}|x||A(x)|)\int_{M}|\nabla_x\Phi(x)|^2dx,\quad\forall \Phi\in W^{1,2}(M)
\end{aligned}\end{equation}
where  $\nu(x)$ is the unit outward normal vector of $\partial M$, and for star shape domain $M$, $x\cdot\nu(x)\geq 0$ ($\forall x\in\partial M$). Particularly, P.B.C. and $|\nabla_x\Phi(x)|^2\leq 2|\partial_\nu\Phi(x)|^2$($\forall x\in\partial M$) as well as the symmetry of $M$ under the transformation $x_j \to -x_j$ ($\forall j=1,2,...,N$) implies the positivity of the second integration in the first line of (\ref{10.28}). From (\ref{10.28}) we find that the only lowest energy state is a constant  with zero eigenvalue in case (3). The proof of (2) is similar.

(4)  Multiplying $H_0\Phi=\lambda \Phi$ by $\exp(2V)\bar\Phi$ and integrating over $M$, in light of $A=\nabla_x V$, we find
\begin{equation}\label{10.29}\begin{aligned}
&\frac12\int_{M}\exp(2V)|\nabla_x\Phi(x)|^2dx=\lambda \int_{M}\exp(2V)|\Phi(x)|^2dx.
\end{aligned}\end{equation}
Then the eigenvalues must be real and the only lowest energy state is a constant with zero eigenvalue. Particularly, for D.B.C., the lowest energy is positive.
 
\vspace{0.5cm}
Suppose $\lambda_0^{H_0}$ and $\lambda_1^{H_0}$ are the 1st and 2nd eigenvalues of $H_0$. Then from Lemma \ref{L3.5},
$$
\lambda_0^{H_0}=0,\quad \lambda_1^{H_0}> 0.
$$\vspace{0.5cm}
We have

\begin{thm}\label{T3.6} Suppose $div_xA(x)\neq$constant and  either (1)
\begin{equation}\label{10.30}\begin{aligned}
\sup_{x\in M}|div A(x)|< \min_{\zeta\in\mathbb{C}:|\zeta|=\frac{|\lambda_1^{H_0}|}2}\frac1{2(1+|\zeta|^2)(1+\|(H_0-\zeta)^{-1}\|^2)^{\frac12}     }
\end{aligned}\end{equation}
as well as  Lemma \ref{L3.5}(3);\\
or (2)  there is $V$ such that $A=\nabla_x V$,   $\Phi$ and $V$ satisfy P.B.C. or $\Phi$ and $V$ satisfy N.B.C..\\
Then  the corresponding ground state of $H$  must be non-constant and $H$ is BRST-supersymmetric breaking in gauge theory provided $H$ is 2-pseudo-supersymmetric. Moreover $H$ is BRST-supersymmetric in W-TFT in case (2).
\end{thm}

$Proof.$ Step 1. Suppose (\ref{10.30}), in light of Lemma \ref{L3.5} and \cite{Kato}(Th.3.18), the first eigenvalue (lowest energy) $\lambda_0^H$ of $H$ is in a small neighborhood of $\lambda_0^{H_0}$ which has the same multiplicity as $\lambda_0^{H_0}$, that is, $\lambda_0^H$ is also simple.  Furthermore the corresponding ground state $\Phi_0^H$ must be non-constant. If not, $\Phi_0^H\equiv C (\neq 0)$, then
\begin{equation}\nonumber\begin{aligned}
H\Phi_0^H=-Cdiv_xA(x)=\lambda_0^HC\Rightarrow div_xA(x)\equiv\lambda_0^H
\end{aligned}\end{equation}
which is contradiction with the assumption $div_xA(x)\neq$constant.

Step 2. Notice that  $\mathcal{Q}\Phi_0^H\neq 0$ for any non-constant state $\Phi_0^H$. Thus there does not exist any physical state in the sense of  gauge theory (see \cite{BBRT}section 3.6). Then $H$ is BRST-supersymmetric breaking in gauge theory  provided $H$ is 2-pseudo-supersymmetric.

Step 3. Suppose there is $V$ such that $A=\nabla_x V$.  Multiplying $H\Phi=\lambda \Phi$ by $\exp(2V)\bar\Phi$ and integrating over $M$, in light of $A=\nabla_x V$ and $\int_{\partial M} e^{2V}\nu(x)\cdot \nabla_x V(x)|\Phi(x)|^2d\mathcal{H}^{N-1}\leq 0$, we find
\begin{equation}\label{10.33}\begin{aligned}
&0\leq \int_{M}\exp(2V)\left(\frac12|\nabla_x\Phi(x)|^2+2|\nabla_xV|^2|\Phi|^2+2Re\bar\Phi\nabla_xV\cdot\nabla_x\Phi  \right)dx=\lambda \int_{M}\exp(2V)|\Phi(x)|^2dx.
\end{aligned}\end{equation}
On the other hand, it is easy to check that $\bar{\mathcal{Q}}\exp(-2V(x))\prod_{j=1}^N\psi^j=0$ and ${\mathcal{Q}}\exp(-2V(x))\prod_{j=1}^N\psi^j=0$. So the lowest energy of $H$ is zero, and $H$ is BRST-supersymmetric in W-TFT.

\vspace{0.5cm}

Remark. 1. Since any $A\in L^2$ can be decomposed as $A(x)=\nabla_x V(x)+ \tilde{A}(x)$ with $div_x\tilde{A}(x)=0$, the condition for $div_x A$ is replaced for $\Delta_x V$.

Remark. 2. Generally, $H$ may be non-Hermitian.  The eigenvalues of $H$ could be negative and complex.    To extend the Witten index to these $H$ (see, for example, \cite{Mostafaza}), the existence and multiplicity of zero eigenvalue to $H$ is important. We have to investigate the condition when  the eigenvalues and their multiplicity of $H$ deviate from the zero eigenvalue of $H_0$. The results in this section mentioned several necessary conditions on the variation of negativity and multiplicity of eigenvalues from $H_0$ to $H$.

\vspace{0.5cm}

\subsection{topological invariant for general pseudo-Hermitian Hamiltonians   }

 As the eigenspaces
of H that are associated with non-zero eigenvalues consist of pseudo-superpartner
pairs of state vectors, under continuous pseudo-supersymmetry preserving
deformations of H or H the Witten index is left invariant. Hence, it is a topological
invariant(see \cite{Mostafaza}). The theory of the topological invariant conditions of the deformations in previous section can be extended to more general pseudo-Hermitian Hamiltonians.
Suppose $H_0$ and $H$ are general pseudo-Hermitian and have pseudo-supersymmetry generated by $\mathcal{Q}_{H_0}$ and  $\mathcal{Q}_{H}$ respectively. Suppose $\lambda_0^{H_0}$ and $\lambda_1^{H_0}$ are the 1st and 2nd eigenvalues of $H_0$, and
$$
\lambda_0^{H_0}=0,\quad \text{Multiplicity of $\lambda_0^{H_0}=1$},\quad  \lambda_1^{H_0}> 0.
$$
Utilizing operator perturbation theory \cite{Kato}(chapt. 3, Th.3.18) we have

\begin{thm}\label{T3.7} Suppose the domain $D(H)\supset D(H_0)$, $\lambda_0^{H_0}$  is simple and
\begin{equation}\label{10.34}\begin{aligned}
\|H-H_0\|< \min_{\zeta\in\mathbb{C}:|\zeta|=\frac{|\lambda_1^{H_0}|}2}\frac1{2(1+|\zeta|^2)(1+\|(H_0-\zeta)^{-1}\|^2)^{\frac12}     }.
\end{aligned}\end{equation}
Then the first eigenvalue $\lambda_0^{H}$ of $H$ is inside the circle $\{\zeta\in\mathbb{C}:|\zeta|<\frac{|\lambda_1^{H_0}|}2\}$, and the multiplicity of $\lambda_0^{H}$  is also simple. That is, if $H$ has zero state then the Witten index  is invariant from $H_0$ to $H$ and if $H$ has not zero state then the $Q$-symmetry of $H$ is breaking.
\end{thm}

\vspace{0.5cm}

As another example, consider an even pseudo-Hermitian Hamiltonian $H$ on Hilbert space $\mathcal{H}=\mathcal{H}_+\bigoplus \mathcal{H}_-$ which has a pseudo-supersymmetry generated by $\mathcal{Q}:\mathcal{H}_\pm\to \mathcal{H}_\mp$. The extended Witten index of pseudo-supersymmetry of $H$ is defined by dim(Ker($H_+$))-dim(ker($H_-$)) where the pseudo-superpartner Hamiltonian $H_\pm:=H\lfloor_{\mathcal{H}_\pm}$. Applying Th.\ref{T3.7} to $H_0=H_+$ and $H=H_-$, we discover that if $H_-$ has zero state then the Witten index  is zero and if $H_-$ has not zero state then there exists a supersymmetric ground state and $H$ is BRST-symmetric.

\vspace{0.5cm}

\section{ bio-conservation law and rhythms arisen in the stochastic neural networks}

In the remaining of this section, we consider (\ref{10.10.0}) with the function
\begin{equation}\label{10.2}
E(u)=\frac{-1}2\sum_{i,j=1}^N w_{ij}u^iu^j-\sum_{i=1}^N\theta_i u^i
\end{equation}
that results in
\begin{equation}\label{10.3}
E^i(u)=-\sum_{j=1}^N \frac{w_{ij}+w_{ji}}{2}u^j-\theta_i,
\end{equation}
and the activation function
\begin{equation}\label{10.4+2}
u^i(t)=F(I_s^i(t)),\quad F(r)=\left\{\begin{aligned}
            &1-e^{-\beta (r_1-r_0)},\quad\forall r\geq r_1\\
            &1-e^{-\beta (r-r_0)},\quad\forall r\in [r_0,r_1]\\
            &0,\quad\forall r<r_0,
            \end{aligned}\right.
\end{equation}
where $\beta>0$, $r_0\in[0,1)$ and $r_1>r_0$. Comparing with the half-wave rectification $[I_s^i(t)-r_0]_+$ (see \cite{Dayan} section 7.2 where $r_0$ is a threshold and the notation $[]_+$ denotes half-wave rectification), as $\beta\to\infty$, this activation function converges to Heaviside function.  Remark that the  conclusions obtained in this section can be easily extended to more general functions $E$ and $F$.

For (\ref{10.4+2}), we have $F''/(F')^2=-1/(1-u^i)$, $F^{-1}(u^i)=\frac1{\beta}\ln\frac1{1-u^i}+r_0$ and $f(u^i)=\beta(1-u^i)$.

\vspace{0.5cm}

\subsection{the quantization of the stochastic neural networks are  BRST supersymmetric  }

In light of (\ref{10.12}) and the topological field theory obtained in the last section, we have

\begin{thm}\label{th3.0} There is $V(x)$ such that $A(x)=\nabla_x V(x)$, $Q$ and $\bar Q$ of (\ref{10.20}) are the BRST operators of the quantization of (\ref{10.10.0}), and the BRST operators of the quantization of (\ref{10.10.0}) is 2-pseudo-supersymmetric(see Corollary \ref{co.10.1}). The quantization of (\ref{10.10.0}) is BRST  supersymmetry  in W-TFT. That is, if the self-organized criticality of the neural networks system can be looked upon as a W-TFT with spontaneously broken BRST symmetry, then the stochastic neural networks (\ref{10.10.0}) which be extensively used in neuroscience \cite{Dayan} are not enough to simulate the SOC.
\end{thm}

\vspace{0.5cm}

\subsection{log spike number vs log window number}

 We solve the stochastic differential equations by numerical method (SDE TooLBOX \cite{Pi}), and try to recover Beggs and Plenz etc.'s results and find power-law.  Fig 1 shows the simulating results of the size distribution of neural activities (the numbers of synchronized firing neurons in a window (1 sec) v.s. the numbers of the window) for seven different weighted connecting matrices $W$ from the stochastic neural networks  with $\beta= 0.1$, $\tau_s=0.001 (=0.01 sec)$, $T = 100$, $N = 20$ (4 inhibited neurons). The weighted connecting matrices $W$ with at least two stable steady states are selected. The distributing curves of log spike number vs log window number is more likely nonlinear than linear.

\begin{figure}[!h]
	\includegraphics[height=8.0cm,angle=0, width=0.8\textwidth]{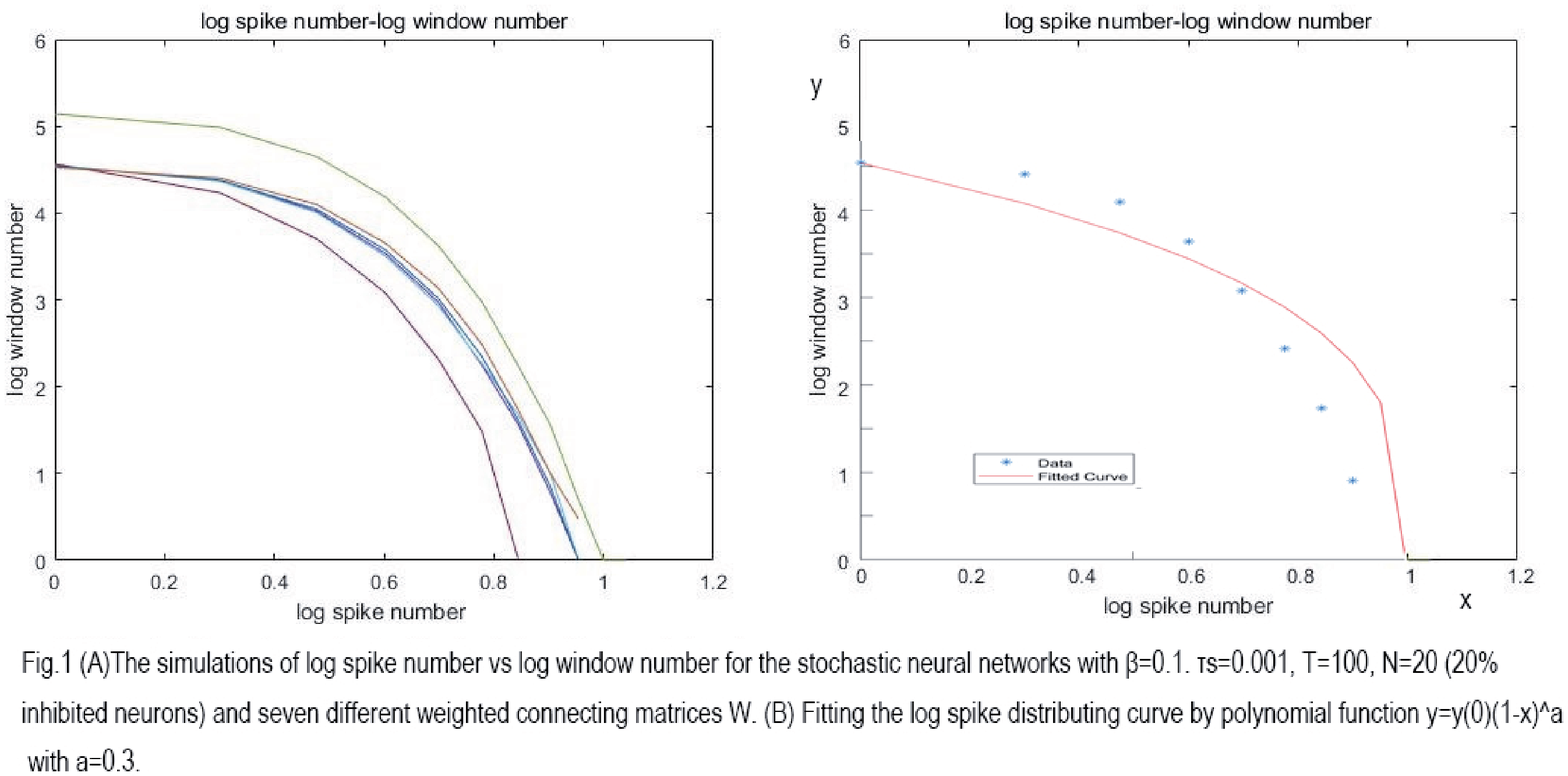}
	\label{fig1}
\end{figure}

\vspace{0.5cm}

\subsection{rhythms arising  from transitions  between up-steady states and down-steady states of the stochastic neural networks}
Even though the stochastic neural networks are supersymmetric in W-TFT and the distributing curves of log spike number vs log window number are more likely non-power-law than power-law, similar rhythms discovered in cortical EEG and ECoG can be simulated by  the stochastic neural networks.

\begin{figure}[!h]
	\includegraphics[height=15.0cm,angle=0, width=0.8\textwidth]{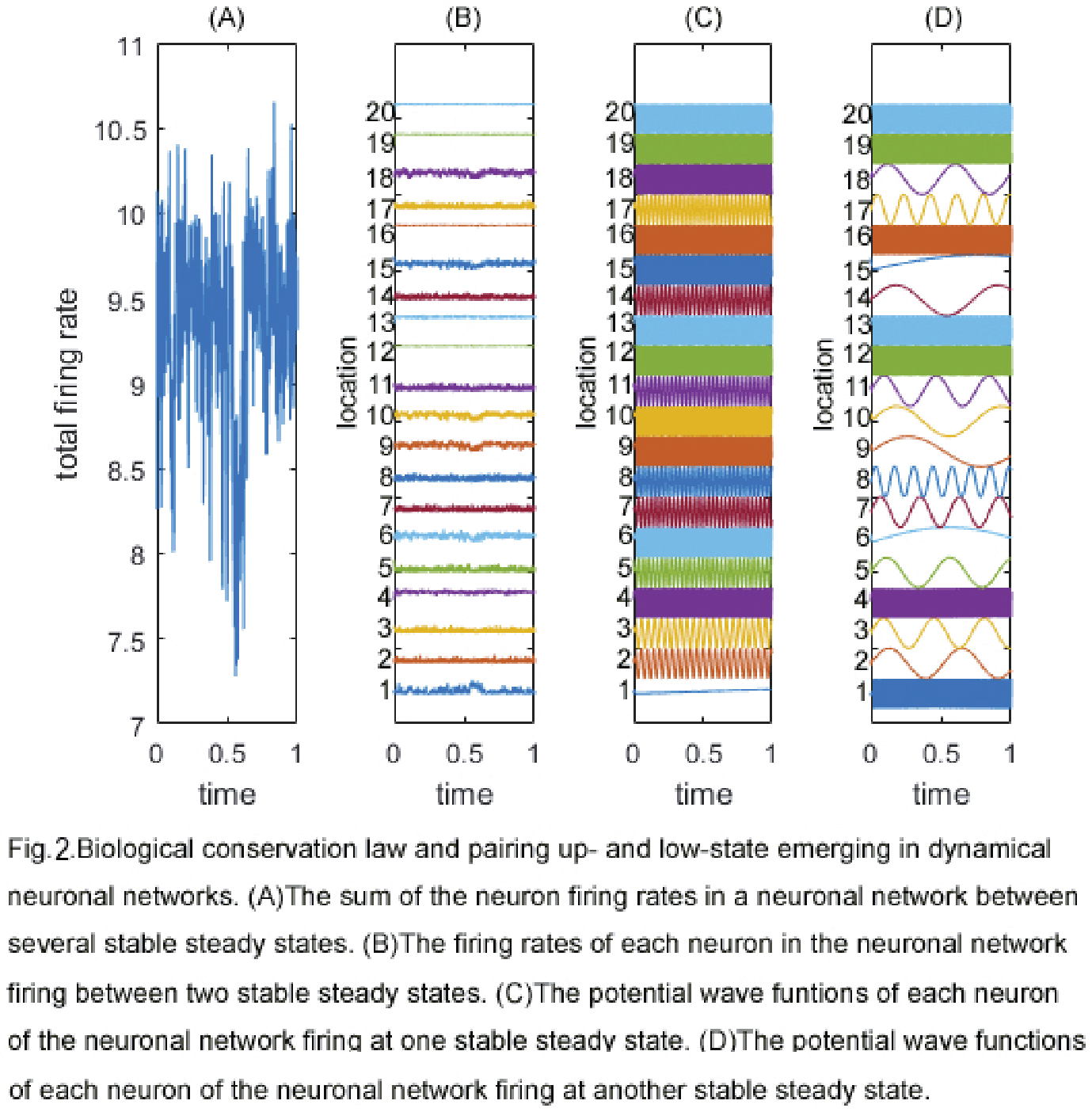}
	\label{fig2}
\end{figure}

Fig 2 is an example  for $N=20$.  Rhythms of the firing rates of the neuronal network  arise from the solution transitioning  between up-steady state $\overline{u}$ and down-steady state $\underline{u}$  of the diffusion equation (\ref{10.6}) (tunneling effect, see Fig 2(A)(B)), meanwhile the biological laws (\ref{10.6.1}) is conserved.

\begin{figure}[!h]
	\includegraphics[height=15.0cm,angle=0, width=0.8\textwidth]{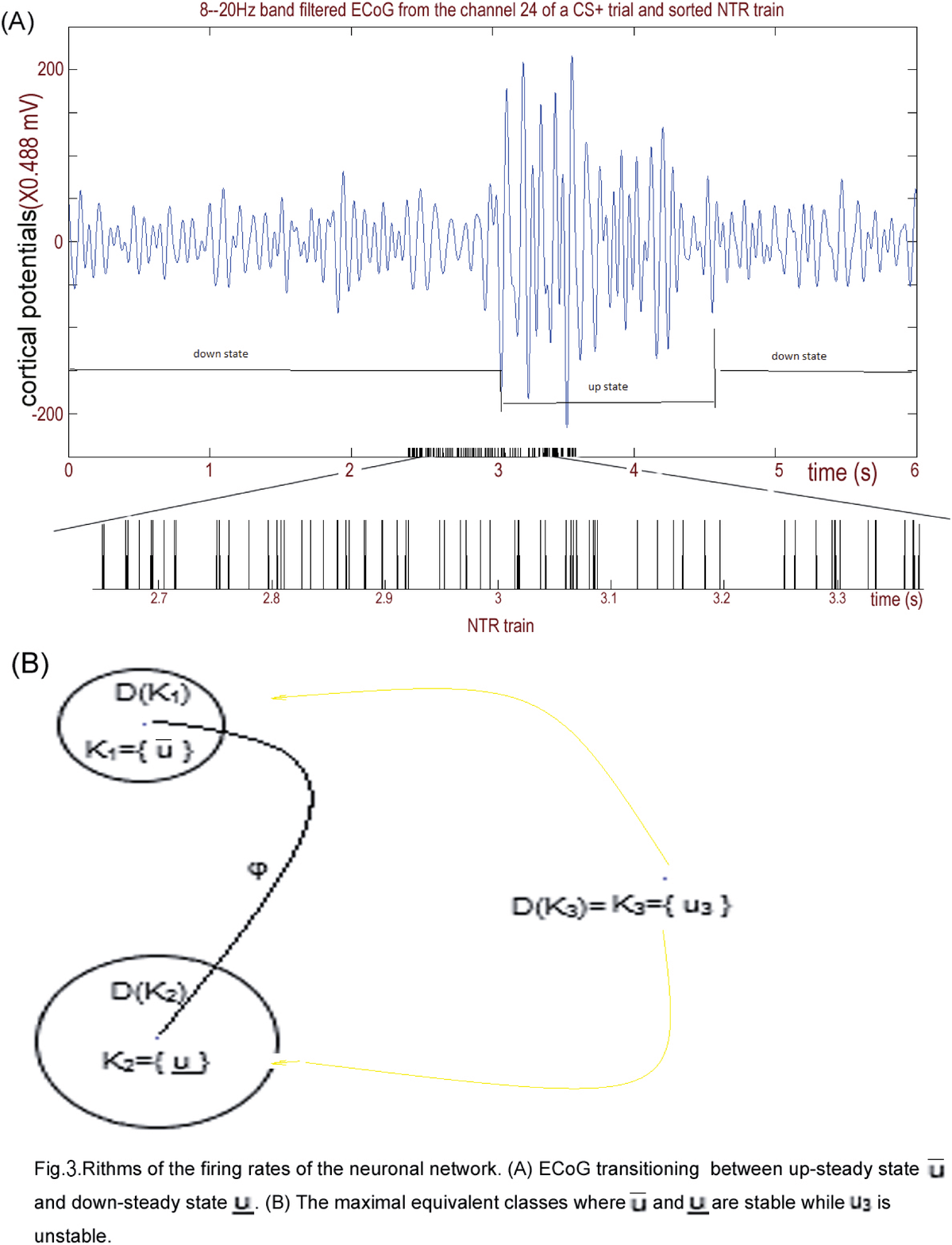}
	\label{fig1}
\end{figure}

\begin{figure}[!h]
	\includegraphics[height=10.0cm,angle=0, width=0.8\textwidth]{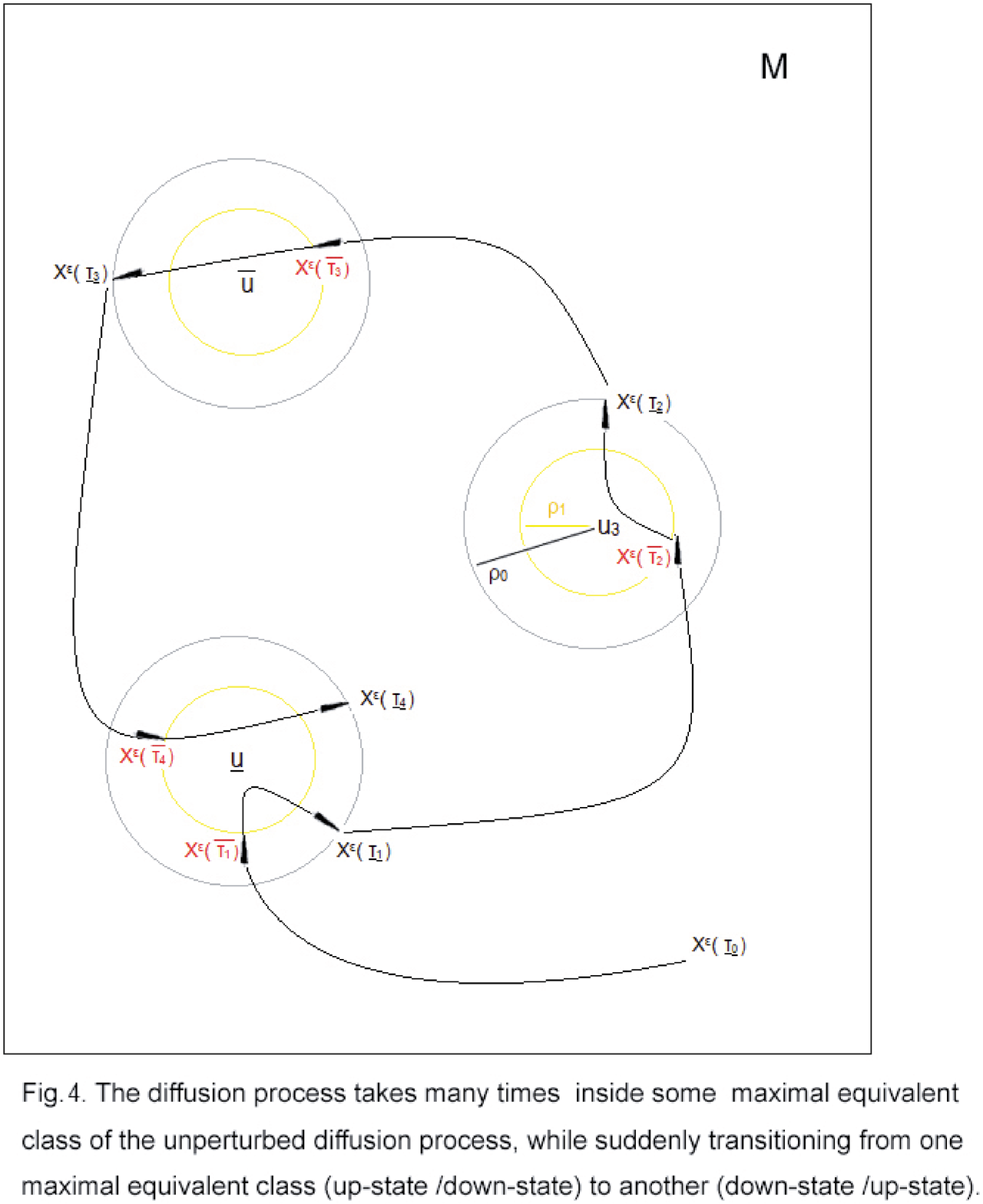}
	\label{fig1}
\end{figure}

Assume the target manifold $M$ is a domain with smooth and compact closure in the state space of the diffusion process (\ref{10.6}).
Suppose $K$  be the maximal equivalent class of a $\omega$-limit set of the unperturbed diffusion process $X(t)$. That is, the set containing at least one $\omega$-limit set,  all whole trajectories of the dynamical system (\ref{10.6}) with $T=0$ starting from any $x_0\in K$. The \textbf{attractable
basin} $D(K)$ of $K$ is the set of initial states from which the
unperturbed diffusion process $X(t)$ converge to $K$
with probability one:
$$
D(K):= \{x\in M:\ P\{\exists \tau>0\,\, s.t \,\,X_{t}\in K,\ \forall
t>\tau|X_{0}=x\}=1\}.
$$
Obviously, $K\subset D(K)$ holds (see Fig 3).

Assume  in $M$ there are a finite number of maximal equivalent classes of the unperturbed diffusion process. Denote by $\partial M$ the boundary of $M$. Let $\mathcal{K}:=\{K_n\}_{n=1}^{n_0}$ be all possible  maximal equivalent sets of the unperturbed diffusion process in $M$.
Let $\rho_0$ be a positive number smaller than half of the minimum of the distances between $K_i$ and $K_j$ and between $K_i$ and $\partial M$. Let $0<\rho_1<\rho_0$. We denote by $O_\rho(K)$ the $\rho$-neighborhood of $K$, by $\partial O_\rho(K)$ the boundary of $O_\rho(K)$. We introduce the random times
\begin{equation}\nonumber\begin{aligned}
\overline{\tau_0}=0, \quad \underline{\tau_n}=\inf\left\{t\geq \overline{\tau_n}: X^\epsilon(t)\in \overline{M}\setminus \cup_{j=1}^{n_0}O_{\rho_0}(K_j)\right\},\\
\overline{\tau_{n+1}}=\inf\left\{t\geq \underline{\tau_n}: X^\epsilon(t)\in \partial M\cup\left(\cup_{j=1}^{n_0}\partial O_{\rho_1}(K_j)  \right)\right\}
\end{aligned}\end{equation}
and consider the Markov chain
$$
X^\epsilon(\overline{\tau_n}),\quad\forall n=0,1,2,....
$$
From $n=1$ on, $X^\epsilon(\overline{\tau_n})$ belongs to $\partial M\cup\left(\cup_{j=1}^{n_0}\partial O_{\rho_1}(K_j)  \right) $. As far as the times $\underline{\tau_n}$ are concerned, $X^\epsilon(\underline{\tau_0})$
can be any point of $\overline{M}\setminus \cup_{j=1}^{n_0}O_{\rho_0}(K_j)$; all the following $X^\epsilon(\underline{\tau_n})\in\overline{M}\setminus \cup_{j=1}^{n_0}O_{\rho_1}(K_j)$, until the time of exit of $X^\epsilon(t)$ to $\partial M$, belong to one of the surfaces $\partial O_{\rho_0}(K_j)$. After exit to the boundary $\partial M$, we have
$$
\overline{\tau_n}=\underline{\tau_n}= \overline{\tau_{n+1}}=\underline{\tau_{n+1}}=...
$$
and the chain $\{X^\epsilon(\overline{\tau_n})\}_n$ stops (see Fig 4). The diffusion process (\ref{10.6}) takes many times  inside some  maximal equivalent class of the unperturbed diffusion process, while suddenly transitioning at some $\bar\tau_n$ from one maximal equivalent class (up-state or down-state) to another (down-state or up-state).  Rhythms of the firing rates of the neuronal network  arise from the solution transitioning  between the maximal equivalent classes  meanwhile the biological law (\ref{10.6.1}) is conserved.

\vspace{0.5cm}

\section{conclusion}

In this paper the quantum field theory is applied to study the
phase transitions in large-scale brain activity, and the associated phenomena associated
with critical behavior.

Suppose SOC can be interpreted as Witten-type topological field theory with spontaneously broken BRST symmetry \cite{BBRT}\cite{Ovchi}. Then the stochastic neural networks (\ref{10.10.0}) which be extensively used in neuroscience \cite{Dayan} are not enough to simulate the SOC. The BRST-symmetry breakdown by instantons is  proved in one-dimension (\cite{BBRT}p.170). In multiple dimension case,  the sufficient and (or) necessary conditions when a Hamiltonian related to the diffusion process (\ref{6.2})  is pseudo-Hermitian and pseudo-supersymmetry, as well as the relation between the interaction representations of the BRST operators  and the Hamiltonian are discovered. Some examples on the hopping evolution of instanton and anti-instanton along the magnetic field  $\tilde{A}$ inducing pseudo-supersymmetry and (or) BRST breakdown are given.

 We find that the distributing curves of log spike number vs log window number are more likely non-power-law than power-law, and the stochastic neural networks do not break BRST supersymmetry in W-TFT. Furthermore, the rhythms discovered in cortical EEG and ECoG can be simulated by  the stochastic neural networks. The sufficient condition on diffusion such that there exists a stationary probability distribution for the stochastic neural networks is obtained. The diffusion process (\ref{10.6}) takes a long time  inside some  maximal equivalent class of the unperturbed diffusion process( (\ref{10.6}) without noise), and suddenly departing from one maximal equivalent class (up-state / down-state) and arriving to another (down-state / up-state).  Rhythms of the firing rates of the stochastic neuronal networks  arise from the transitions  between the maximal equivalent classes (tunneling effect),  meanwhile the biological law (\ref{10.6.1}) is conserved.

\vspace{0.5cm}

\section*{Acknowledgments}
We would like to thank Prof. ShiGang Chen for bringing the subject to our attention and for stimulating discussions. J.Zhai and CJ.Yu were supported by National Natural Science Foundation of China (11671354). Y.Zhai was supported by National Natural Science Foundation of China (81801644).

\bibliographystyle{plain}
\bibliography{ref}

\end{document}